\documentclass[aps,showpacs,english,preprint,nofootinbib]{revtex4}
\usepackage{mathrsfs}
\usepackage{graphicx}
\usepackage{amsmath, amssymb}
\usepackage{babel}
\usepackage{color}

\begin{document}

\preprint{ACFI-T14-15}


\title{Reexamination of The Standard Model Nucleon Electric Dipole Moment}

\author{Chien-Yeah Seng$^{a}$}

\affiliation{$^{a}$Amherst Center for Fundamental Interactions\\
Department of Physics, University of Massachusetts Amherst\\
Amherst, MA 01003 USA}

\date{\today}

\begin{abstract}

The Cabibbo-Kobayashi-Maskawa matrix in the Standard Model is
currently the only experimentally-confirmed source of CP-violation.
The intrinsic electric dipole moment of the nucleon induced by this
CP-phase via hadronic loop and pole diagrams was studied more than
two decades ago, but is subject to various theoretical issues such
as the breakdown of chiral power counting and uncertainties in the
determination of low energy constants. I carry out an up-to-date
re-analysis on both one-loop and pole diagram contributions to the
nucleon electric dipole moment based on Heavy Baryon Chiral
Perturbation Theory in a way that preserves power counting, and
Iredo the determination of the low energy constants following the
results of more recent articles. Combined with an estimation of
higher-order contributions, I expect the long-distance contribution
to the Standard Model nucleon electric dipole moment to be
approximately $(1\times10^{-32}-6\times10^{-32})e\,\mathrm{cm}$.

\end{abstract}


\maketitle

\section{Introduction}

The search for permanent electric dipole moment (EDM) of elementary
and composite particles is motivated by its CP-violating nature. We
live in a universe in which the amount of baryons and antibaryons
are unequal. In order to explain this asymmetry CP-violating
interactions are needed to fulfill one of the three Sakharov
criteria \cite{Sakharov:1967dj}. EDMs of elementary and composite
particles are, in most cases, direct consequences of these
interactions which can be probed in low-energy experiments. Since
the first upper limit on the neutron EDM obtained by Smith, Purcell
and Ramsey in 1957 \cite{Smith:1957ht}, numerous experiments have
been performed to improve the sensitivity of EDM measurements in
different particle systems. Currently, the most stringent bounds on
EDMs are set for the electron ($8.7\times10^{-29}e\,\mathrm{cm}$,
90\% C.L.)\cite{Baron:2013eja} and the mercury atom
($3.1\times10^{-29}e\,\mathrm{cm}$, 95\%
C.L.)\cite{Griffith:2009zz}, while the current upper limit on
neutron and proton EDMs are $2.9\times10^{-26}e\,\mathrm{cm}$ (90\%
C.L.)\cite{Baker:2006ts} and $7.9\times10^{-25}e\,\mathrm{cm}$ (95\%
C.L.) respectively (the latter is deduced from the bound on the
mercury EDM). Future experiments are designed (or have been
considered) to push these bounds even further down. For the neutron
EDM, this includes the experiment at Paul Scherrer Institut
(PSI)\cite{Altarev:2009zz}, the CryoEDM and PNPI/ILL experiment at
Institut Laue-Langevin (ILL)\cite{Serebrov:2009}, the SNS neutron
EDM experiment at Oak Ridge, the TRIUMF experiment in Canada and the
Munich experiment at Germany. These experiments are designed to
reach a $10^{-28}e\,\mathrm{cm}$ precision level for the neutron EDM
\cite{Kumar:2013qya}. Also, both COSY\cite{Lehrach:2012eg} and
BNL\cite{Semertzidis:2011qv} have proposed storage ring experiments
designed to measure the proton EDM to a level of
$10^{-29}e\,\mathrm{cm}$ precision.

Although numerous Beyond Standard Model (BSM) scenarios have been
proposed that give rise to measurable EDMs within current
experimental precision level, so far no definitive signal of such
physics has been observed \footnote{There are indeed some hopeful
candidates, for example the muon $g-2$ anomaly; but no conclusive
statement can be made before one could further improve the
experimental precision and reduce the theoretical uncertainty of the
SM prediction.}. Therefore, the CP-violating phase of the
Cabibbo-Kobayashi-Maskawa (CKM) matrix in the Standard Model (SM)
remains the only source for intrinsic EDMs. Questions have been
raised concerning the expected size of EDMs coming from purely SM
physics\cite{Pospelov}. A simple dimensional analysis using
constituent quark masses may suggest that the SM-induced neutron EDM
could be as large as $10^{-29}e\,\mathrm{cm}$, approaching the level
of sensitivity for future EDM experiments. It is therefore important
to have a better estimate for the SM contribution to the nucleon
EDM. To leading order, the quark EDM induced by the CKM matrix
starts at three-loops \cite{Shabalin:1979gh}. A detailed calculation
showed that the valence-quark contribution to the neutron EDM is of
order $10^{-34}e\,\mathrm{cm}$ \cite{Czarnecki:1997bu}. It was also
shown that long-distance contributions, namely contributions with
baryons and mesons as effective degrees of freedom (DOFs), could
generate a much larger hadronic EDM. For instance, the pion-loop
contribution to the neutron EDM was first studied in a paper by
Barton and White \cite{Barton:1969gi} which produced log-divergent
results in the chiral limit indicating that the long-range
contribution may dominate. On the other hand, in a series of papers,
Gavela et.al. studied the pole-diagram contribution with the
CP-violating phase generated by $|\Delta S|=1$ electroweak
\cite{Gavela:1981sk} and gluonic penguin diagrams
\cite{Gavela:1981sm}. They claimed that the latter is dominant and
derived a SM neutron EDM of order $10^{-31}e\,\mathrm{cm}$. The
possibility of a long-range contribution to the neutron EDM from the
CKM matrix was first pointed out by Khriplovich and
Zhitnitsky\cite{Khriplovich:1981ca}. He et.al \cite{He:1989xj} did a
thorough chiral-loop calculation and re-analyzed the pole-diagram
contribution in \cite{Gavela:1981sk,Gavela:1981sm} and argued that
the two are of the same order of magnitude. Their estimate for the
neutron EDM is
$1.6\times10^{-31}e\,\mathrm{cm}-1.4\times10^{-33}e\,\mathrm{cm}$,
which is currently the most widely accepted estimate for the SM
neutron EDM. In recent years, the charm contribution to nucleon EDMs
is also considered and it is roughly $10^{-31}e$ cm
\cite{Mannel:2012qk}.

The purpose of this paper is to revisit the previous study of both
chiral-loop and the pole contributions to the nucleon EDM in order
to sharpen our SM benchmark value. On the theoretical side, one
could improve earlier work in several ways. For instance, the chiral
loop calculation in \cite{He:1989xj} adopted an older meson theory
utilizing a pseudoscalar strong baryon-meson coupling that should be
replaced by the standard axial-vector coupling. Also, their work
that utilized an effective hadronic Lagrangian in computing
chiral-loop diagrams faced another well-known problem in the loss of
power counting similar to that happening in the relativistic Chiral
Perturbation Theory (ChPT). ChPT is a non-renormalizable theory that
involves infinitely many interaction terms. Its predictive power
therefore relies on the fact that higher order terms are suppressed
by powers of $p/\Lambda_\chi$ where $p$ is the typical mass or
momentum scales of hadronic DOFs and $\Lambda_\chi\sim 1$GeV. This
expansion however becomes ambiguous when baryons are included
because a typical baryon mass is $M_B\sim 1$GeV. Therefore,
$M_B/\Lambda_\chi$ is no longer a small expansion parameter. Heavy
Baryon Chiral Perturbation Theory (HBchPT) \cite{Jenkins:1990jv}
provides a convincing way to get around this issue by performing a
field redefinition in the baryon field to scale out its
mass-dependence. By doing this, one can split the baryon field into
``light" and ``heavy" components, where the former depends only on
its residual momentum which is well below 1 GeV. After integrating
out the heavy component of the baryon field, the effective
Lagrangian can be written as a series expansion of $1/m_N$. This
eliminates the possibility of a factor $m_N$ appearing in the
numerator and thus restores the power counting. Many works have
appeared recently calculating the nucleon EDM induced by different
BSM physics using HBchPT (see \cite{Engel:2013lsa} for a general
overview). Although the convergence of the SU(3) HBchPT is not as
good as its SU(2) counterpart because $m_K/m_N$ is not very small
\cite{Jenkins:1992pi,Meissner:1997hn,Durand:1997ya,Puglia:1999th,Puglia:2000jy},
it is still theoretically beneficial as it provides a formal
classification of different contributions into leading and
sub-leading orders. In this work, the chiral-loop contribution to
the nucleon EDM are recalculated up to the leading-order (LO) in the
heavy baryon (HB)-expansion.

Additionally, previous numerical results of loop and pole
contributions face large uncertainties due to poorly-known values of
physical constants in the weak sector at that time. For example, the
CP-violating phase $\delta$ of the CKM matrix quoted in Ref.
\cite{He:1989xj} had an uncertainty that spans one order of
magnitude. The fitting of certain low energy constants (LECs) such
as weak baryon-meson interaction strengths, has been updated since.
Also, their theoretical estimation of various CP-phases in the
effective weak Lagrangian was based on older work
\cite{Guberina:1979ix,Donoghue:1986hh} which had been improved by
others. Furthermore, for previous work on pole contributions, their
estimation on effective CP-phases was based only on a single gluonic
penguin operator without considering the full analysis of operator
mixing and renormalization group running. Moreover, the approximate
form of their analytic expressions was based on the out-of-date
assumption that $m_t\ll m_W$. In this work, I do a more careful
determination of weak LECs, taking all these issues into account.
Combining my calculation and an estimate of higher-order effects, I
predict a range of the long-distance SM contribution to the nucleon
EDM to be around $(1-6)\times10^{-32}e\,\mathrm{cm}$. I identify the
main sources of uncertainty and discuss possible steps one could
take to improve upon that. At the same time, I use dimensional
analysis to estimate the size of possible short-distance
counterterms. I find that they could be as large as
$4\times10^{-32}e\,\mathrm{cm}$.

This work is arranged as follows: in Section II I will briefly
outline the main ingredients of the SU(3) HBchPT and introduce the
weak Lagrangian responsible for the generation of the nucleon EDM.
In Section III I will determine the LECs. In Section IV and V
Iderive the analytic expressions for loop and pole contributions to
the nucleon EDM respectively and calculate their numerical values.
In Section VI I will provide some further discussions and draw my
conclusions.

\section{HBchPT: Strong and Electroweak Interactions}

In this section, I review some basic concepts of ChPT with the
primary aim of establishing conventions and notation. ChPT is a
low-energy effective field theory (EFT) of quantum chromodynamics
(QCD) with hadrons as low energy DOFs. QCD exhibits a global chiral
symmetry in the limit of massless quarks. However this symmetry is
spontaneously broken in the ground state and leads to the emergence
of Goldstone bosons which are identified as pseudoscalar mesons. The
corresponding EFT obeys the same symmetry. An infinite tower of
operators respecting the symmetry with increasing mass dimensions is
organized in the Lagrangian. However, only a finite number of
operators are retained since the the dropped higher-dimensional
operators make contributions that are suppressed by powers of
$p/\Lambda_\chi$.

I use the standard non-linear representation of chiral fields
\cite{Scherer:2002tk,Donoghue:1992dd,Georgi:1985kw}, in which the
pseudoscalar meson octet is incorporated in the exponential function
$U=\mathrm{exp}\{i\phi/F_\pi\}$, where
\begin{equation}
\phi=\sum_{a=1}^{8}\phi_a\lambda_a=\left(\begin{array}{ccc}
\pi^0+\frac{1}{\sqrt{3}}\eta_8 & \sqrt{2}\pi^+ & \sqrt{2}K^+\\
\sqrt{2}\pi^- & -\pi^0+\frac{1}{\sqrt{3}}\eta_8 & \sqrt{2}K^0\\
\sqrt{2}K^- & \sqrt{2}\bar{K}^0 &
-\frac{2}{3}\eta_8\end{array}\right) \end{equation}with
$F_\pi\approx 93$MeV. The matrix $U$ transforms under the chiral
rotation as: $U\rightarrow LUR^\dagger$, where $L$ and $R$ are
elements of $SU(3)_L$ and $SU(3)_R$ respectively. The mass term of
the meson octet is introduced using spurion analysis: the QCD
Lagrangian would exhibit chiral invariance if the quark mass matrix
$M=\mathrm{diag}\{m_u,m_d,m_s\}$ transforms as $M\rightarrow
LMR^\dagger$. Therefore, its low energy effective theory written in
terms of the spurion field $M$ should also exhibit a similar
invariance. The lowest-order operator that is invariant is
$\mathrm{Tr}[MU^\dagger+UM^\dagger]$. This operator gives rise to
non-zero meson masses which are isospin-symmetric.

The ground state $J^P=(1/2)^+$ baryon octet is assembled into the
matrix:
\begin{equation}
B=\left(\begin{array}{ccc}
\frac{\Sigma^0}{\sqrt{2}}+\frac{\Lambda}{\sqrt{6}} & \Sigma^+ & p\\
\Sigma^- & -\frac{\Sigma^0}{\sqrt{2}}+\frac{\Lambda}{\sqrt{6}} & n\\
\Xi^- & \Xi^0 & -\frac{2\Lambda}{\sqrt{6}}\end{array}\right).
\end{equation} It transforms as: $B\rightarrow
KBK^\dagger$ with $K=K(L,R,U)$ being a unitary matrix. In order to
couple baryons with the pseudoscalar octet, we define $\xi=\sqrt{U}$
which transforms as $\xi\rightarrow L\xi K^\dagger=K \xi R^\dagger$
and introduce the Hermitian axial vector:
\begin{equation}
\mathcal{A}_\mu=\frac{i}{2}[\xi\partial_\mu\xi^\dagger-\xi^\dagger\partial_\mu
\xi]\end{equation}which transforms as $\mathcal{A}_\mu\rightarrow
K\mathcal{A}_\mu K^\dagger$ under the chiral rotation (we have
neglected its coupling with external fields because it is not needed
in this work).

I now proceed with with the formulation of HBchPT. In order to scale
out the heavy mass-dependence, I rewrite its momentum as
\begin{equation}p_\mu=m_N v_\mu+k_\mu,\end{equation}
where $m_N$ is the nucleon mass, $v_\mu$ is the velocity of the
baryon (which is conserved in the $m_N\rightarrow \infty$ limit) and
$k_\mu$ is the residual momentum of the baryon which is well below
$1$ GeV. I therefore rescale the baryon field and retain its
``light" component \footnote{in the sense that it only depends on
the residual momentum}:
\begin{equation}
B_v(x)=e^{im_N v\cdot x}\frac{1+v\!\!\!/}{2}B(x)\end{equation} The
subscript $v$ will be dropped from now on. OI integrate out the
remaining component which is ``heavy". The baryon propagator thus
becomes:
\begin{equation}iS_B(k)=\frac{i}{v\cdot k-\delta_B+i\epsilon}\end{equation}where $\delta_B=m_B-m_N$ is
the baryon mass splitting. This procedure also reduces Dirac
structures to either 1 or $S^\mu$ with the latter being the
spin-matrix of the baryon satisfying $S\cdot v=0$. In this work I
concentrate only on terms that are leading order in the HB-expansion
(with the exception of the baryon electromagnetic dipole transition
operator that appears in pole diagrams as I will explain below).

The lowest-order strong Lagrangian involving only the $(1/2)^+$
baryons, Goldstone bosons and electromagnetic fields relevant to our
work is given by:

\begin{eqnarray}\label{eq:CP_even_Lagrangian}
\mathcal{L}&=&\frac{F_\pi^2}{4}\mathrm{Tr}[\mathcal{D}_\mu
U\mathcal{D}^\mu U^\dagger]+\frac{F_\pi^2}{4}\mathrm{Tr}[\chi_+]+\mathrm{Tr}[\bar{B}iv\cdot\mathcal{D}B]+2D\mathrm{Tr}[\bar{B}S^\mu\{\mathcal{A}_\mu,B\}]+2F\mathrm{Tr}[\bar{B}S^\mu[\mathcal{A}_\mu,B]]\nonumber\\
&&+\frac{b_D}{2B_0}\mathrm{Tr}[\bar{B}\{\chi_+,B\}]+\frac{b_F}{2B_0}\mathrm{Tr}[\bar{B}[\chi_+,B]]+\frac{b_0}{2B_0}\mathrm{Tr}[\bar{B}B]\mathrm{Tr}[\chi_+]\end{eqnarray}
where $D=0.80$, $F=0.50$ \cite{Scherer:2002tk} and $\mathcal{D}_\mu
U=\partial_\mu U+ieA_\mu[Q,U]$. Here
$Q=\mathrm{diag}\{2/3,-1/3,-1/3\}$ is the quark charge matrix while
$B_0$ is a parameter characterizing the chiral quark condensate and
$\chi_+=2B_0(\xi^\dagger M\xi^\dagger+\xi M\xi)$ introduces the
quark-mass dependence. The last three terms in Eq.
\eqref{eq:CP_even_Lagrangian} are responsible for the mass splitting
within the baryon octet \cite{Kaiser:1995eg}. Since I have scaled
out the nucleon mass from the baryon field $B$ the proton and
neutron will appear as massless excitations and the other baryons
will have an excitation energy given by the ``residual" mass
$\delta_B$. This is important later during the computation of pole
diagrams.

For the purpose of pole diagram contributions I need also to include
the $(1/2)^-$ baryon octet. The importance of these resonances can
be traced back to the observation of the unexpectedly large
violation of Hara's theorem \cite{Hara:1964zz} which states that the
parity-violating radiative $B\rightarrow B'\gamma$ transition
amplitude should vanish in the exact SU(3) limit. The authors of
Ref. \cite{LeYaouanc:1978ef} (and later improved by
\cite{Borasoy:1999nt}) pointed out that this apparent puzzle could
be resolved by including baryon resonances that give rise to pole
diagrams which enhance the violation of Hara's theorem. Therefore,
one should naturally expect that the same kind of diagrams will also
play an important role in the determination of the nucleon EDM. The
resonance $(1/2)^-$ octet is denoted as $\mathcal{R}$:
\begin{equation}
\mathcal{R}=\left(\begin{array}{ccc}
\frac{\Sigma^{0*}}{\sqrt{2}}+\frac{\Lambda^*}{\sqrt{6}} & \Sigma^{+*} & p^*\\
\Sigma^{-*} & -\frac{\Sigma^{0*}}{\sqrt{2}}+\frac{\Lambda^*}{\sqrt{6}} & n^*\\
\Xi^{-*} & \Xi^{0*} &
-\frac{2\Lambda^*}{\sqrt{6}}\end{array}\right).\end{equation}It
transforms in the same way as $B$ except that it has a negative
intrinsic parity.

The part of strong and electromagnetic chiral Lagrangian involving
$\mathcal{R}$ which is relevant to our work is given by:
\begin{eqnarray}
\mathcal{L}_\mathcal{R}&=&\mathrm{Tr}[\bar{\mathcal{R}}iv\cdot\mathcal{D}\mathcal{R}]-\bar{\delta}_{\mathcal{R}}\mathrm{Tr}[\bar{\mathcal{R}}\mathcal{R}]+\frac{\tilde{b}_D}{2B_0}\mathrm{Tr}[\bar{\mathcal{R}}\{\chi_+,\mathcal{R}\}]+\frac{\tilde{b}_F}{2B_0}\mathrm{Tr}[\bar{\mathcal{R}}[\chi_+,\mathcal{R}]]+\frac{\tilde{b}_0}{2B_0}\mathrm{Tr}[\bar{\mathcal{R}}\mathcal{R}]\mathrm{Tr}[\chi_+]\nonumber\\
&&-2r_D(\mathrm{Tr}[\bar{\mathcal{R}}(v_\mu S_\nu-v_\nu
S_\mu)\{f^{\mu\nu}_+,B\}]+\mathrm{Tr}[\bar{B}(v_\mu S_\nu-v_\nu
S_\mu)\{f^{\mu\nu}_+,R\}])\nonumber\\
&&-2r_F(\mathrm{Tr}[\bar{\mathcal{R}}(v_\mu S_\nu-v_\nu
S_\mu)[f^{\mu\nu}_+,B]]+\mathrm{Tr}[\bar{B}(v_\mu S_\nu-v_\nu
S_\mu)[f^{\mu\nu}_+,\mathcal{R}]]).\end{eqnarray} The second to
fifth terms of $\mathcal{L}_\mathcal{R}$ give the average residual
mass and mass-splitting among the $(1/2)^-$ baryon octet. Constants
$r_D$ and $r_F$ are electromagnetic coupling strengths between $B$
and $\mathcal{R}$ and $f^{\mu\nu}_+$ is the chiral field strength
tensor of the electromagnetic field that, in the SU(3) version of
ChPT, is given by \cite{Scherer:2002tk}:
\begin{equation}f^{\mu\nu}_+=-e[\xi^\dagger Q\xi+\xi
Q\xi^\dagger]F^{\mu\nu}\end{equation}with $e>0$. The reason we
include $r_D$ and $r_F$ terms even though they are formally
$1/m_N$-suppressed is that they will then be compensated by small
denominator $\delta_B$ factors in pole diagrams.

Next I introduce the relevant weak Lagrangian that gives rise to the
nucleon EDM. As the only CP-violating effect in the SM is the
complex phase in the CKM matrix, the strange quark must be included.
The CP-phase is attached to various $|\Delta S|=1$ four-quark
operators that are responsible for kaon decay and nonleptonic
hyperon decays. It is well-known that the product of two charged
weak currents could transform as $(8_L,1_R)$ or $(27_L,1_R)$ under
the SU(3) chiral rotation. Extra $|\Delta S|=1$ operators could be
induced via gluonic or electroweak penguin diagrams. The former
transforms as $(8_L,1_R)$ while the latter may introduce a
$(8_L,8_R)$ component that is however suppressed by the smallness of
the fine structure constant. Furthermore, since $(8_L,1_R)$
operators have isospin $I=1/2$ while $(27_L,1_R)$ operators can have
both $I=1/2$ and $I=3/2$ components we would naturally expect the
latter to be subdominant as compared to the $(8_L,1_R)$ operators.
Otherwise the $I=3/2$ channel would be as important as the $I=1/2$
channel in non-leptonic decay processes, violating the
experimentally observed $|\Delta I|=1/2$ dominance in these
processes. Hence, effective operators I introduce later should also
transform as $(8_L,1_R)$.

The pure mesonic Lagrangian that triggers the $|\Delta I|=1/2$ kaon
decay channel is given by \cite{Donoghue:1992dd}:
\begin{equation}
\mathcal{L}_8=g_8 e^{i\varphi}\mathrm{Tr}[\lambda_+ D_\mu U D^\mu
U^\dagger]+h.c
\end{equation}
where $\lambda_+=(\lambda_6+i\lambda_7)/2$. The non-zero value of
$\varphi$ introduces the CP-violating effect. Meanwhile, the
corresponding baryonic operator that triggers the nonleptonic
hyperon decay is given by \cite{Tandean:2002vy}:
\begin{equation}
\mathcal{L}_w^{(s)}=h_D
e^{i\varphi_D}\mathrm{Tr}[\bar{B}\{\xi^\dagger\lambda_+\xi,B\}]+h_Fe^{i\varphi_F}\mathrm{Tr}[\bar{B}[\xi^\dagger\lambda_+\xi,B]]+h.c.
\end{equation}
Here the superscript $(s)$ indicates that these operators mediate
S-wave decays. In principle there is a counterpart operator with the
Dirac structure $\gamma_5$, which is time-reversal odd and is
proportional to the complex phase in the CKM matrix. I do not need
this extra operator because it vanishes at leading order in the
HB-expansion upon the non-relativistic reduction of the Dirac
structure. Also, our definitions of $h_D$ and $h_F$ here are
slightly different from \cite{Tandean:2002vy} as we take $h_D,h_F$
to be real, with the complex phases explicitly factored out.

Finally, for the purpose of including pole-diagram contributions, I
need the weak Lagrangian that triggers the $B-\mathcal{R}$
transition. The lowest order Lagrangian is given by
\cite{Borasoy:1999md}:

\begin{equation}
\mathcal{L}_w^{B\mathcal{R}}=iw_D
e^{i\tilde{\varphi}_D}\mathrm{Tr}[\bar{\mathcal{R}}\{
h_+,B\}]+iw_Fe^{i\tilde{\varphi}_F}\mathrm{Tr}[\bar{\mathcal{R}}[
h_+,B]]+h.c
\end{equation}
where $h_+\equiv\xi^\dagger\lambda_+\xi+\xi^\dagger\lambda_-\xi$.
The counterpart with a $\gamma_5$ structure similarly vanishes at
leading order in the HB-expansion.

\section{Determination of the LECs}

There are altogether 12 LECs that enter into the estimate for the
nucleon EDM: seven interaction strengths
$\{r_D,r_F,g_8,h_D,h_F,w_D,w_F\}$ and five CP-violating phases
$\{\varphi,\varphi_D,\varphi_F,\tilde{\varphi}_D,\tilde{\varphi}_F\}$.
They are either extracted from experiments or obtained by
theoretical modeling \footnote{Unfortunately, none of these LECs in
the literature come with error bars, so I cannot estimate the error
introduced by the fitting of LECs.}.

Pure electromagnetic $B-\mathcal{R}$ transition coupling strengths
$r_D$ and $r_F$ are fitted to electromagnetic decays of $(1/2)^-$
resonances. The authors of Ref. \cite{Borasoy:1999nt} obtain:
\begin{equation}er_D=0.033\mathrm{GeV}^{-1},
er_F=-0.046\mathrm{GeV}^{-1}.\end{equation}

The constant $g_8$ is fitted to the $K_s^0\rightarrow \pi^+\pi^-$
decay rate, ignoring the small CP-violating effect
\cite{Beringer:1900zz}, giving
\begin{equation}g_8=6.84\times10^{-10}\mathrm{GeV}^2.\end{equation}

The CP-phase $\varphi$ is, up to a negative sign, the phase of the
$K^0\rightarrow \pi\pi(I=0)$ decay amplitude:
\begin{equation}
\varphi=-\xi_0=-\frac{\mathrm{Im}A_0}{\mathrm{Re}A_0}\end{equation}
In principle one could extract $\xi_0$ from the measurement of the
CP-violating parameter $\epsilon'$ in the kaon decay. However,
$\epsilon'$ is a linear combination of $\xi_0$ and another
CP-violating phase, $\xi_2$, of the $I=3/2$ channel. Simple
estimation \cite{Donoghue:1992dd} suggests that $\xi_2$ is of the
same order as $\xi_0$ making $\xi_0$ hard to extract directly from
the experiment. I therefore refer to theoretical estimation based on
the large-$N_c$ approach \cite{Buras:2008nn} which gives:
\begin{equation}\varphi=-\xi_0\approx-\sqrt{2}|\epsilon|\times(-6\times10^{-2})\approx1.89\times10^{-4}\approx6.4J,\end{equation}
where $J=(2.96^{+0.20}_{-0.16})\times10^{-5}$ \cite{Beringer:1900zz}
is the Jarlskog invariant\cite{Jarlskog:1985ht}. It is worthwhile to
mention that, in Ref. \cite{He:1989xj} the uncertainty of $J$ spans
an order of magnitude leading to the main source of the error in the
estimate of the neutron EDM during that time. Today, $J$ is
determined with much higher precision so the associated uncertainty
is sub-leading compared to uncertainties due to higher-order effects
in the HB-expansion and unknown short-distance counterterms, which
we will discuss later.

The four remaining interaction strengths $h_D,h_F,w_D,w_F$ were
determined in \cite{Borasoy:1999md} by simultaneously fitting them
to the s and p-wave amplitudes of nonleptonic hyperon
decays:\begin{equation} h_D\approx
0.44\times10^{-7}\mathrm{GeV},h_F\approx
-0.50\times10^{-7}\mathrm{GeV},w_D\approx-1.8\times10^{-7}\mathrm{GeV},w_F\approx2.3\times10^{-7}\mathrm{GeV}.\end{equation}
The last two constants were determined by setting
$m_\mathcal{R}\approx 1535$MeV.

Finally, I need to know the four remaining CP-phases
$\{\varphi_D,\varphi_F,\tilde{\varphi}_D,\tilde{\varphi}_F\}$. These
phases have been considered in ref \cite{Gavela:1981sk}, but their
treatments are less satisfactory due to the neglect of the operator
mixing effect and a certain outdated approximation of the small top
quark mass assumption. In order to improve upon that, I review a
more recent work done in Ref. \cite{Tandean:2002vy} that determined
$\{\varphi_D,\varphi_F\}$ and apply scaling arguments to provide an
estimate of $\{\tilde{\varphi}_D,\tilde{\varphi}_F\}$. Ref.
\cite{Tandean:2002vy} pointed out that after considering operator
mixing and renormalization group running, the dominant operator that
gives rise to the CP-violating phase in the $|\Delta S|=1$, $|\Delta
I|=1/2$ sector is given by:\begin{equation}\hat{Q}_6=-2\sum_q
\bar{d}(1+\gamma_5)q\bar{q}(1-\gamma_5)s.\end{equation} Ref
\cite{Tandean:2002vy} then computed the factorizable and
non-factorizable contributions to $\varphi_D,\varphi_F$ induced by
$\hat{Q}_6$. Here ``factorizable'' means to regard $\hat{Q}_6$ as a
product of two chiral quark densities and match it to chiral
operators. The matching is done by realizing that
$\bar{q}_Rq_L\sim\partial\mathcal{L}_{QCD}/\partial
m_q=\partial\mathcal{L}_{\mathrm{chiral}}/\partial m_q$. On the
other hand, the ``non-factorizable'' contribution is obtained simply
by taking the hadronic matrix element of $\hat{Q}_6$ using the quark
model. These two contributions are distinct because the factorizable
piece contains a factor of chiral quark condensate $F_\pi^2B_0$
through:
\begin{equation}
\left\langle0\right|\bar{q}^i_Lq^j_R\bar{q}^k_Rq^l_L\left|B\bar{B}'\right\rangle\sim
\left\langle0\right|\bar{q}^i_Lq^j_R\left|0\right\rangle
\left\langle0\right|\bar{q}^k_Rq^l_L\left|B\bar{B}'\right\rangle=-\frac{1}{2}F_\pi^2B_0\delta_{ij}\left\langle0\right|\bar{q}^k_Rq^l_L\left|B\bar{B}'\right\rangle
\end{equation}
while the same quantity never appears in a quark model calculation.
Combining the two, they found $\mathrm{Im}(h_D\exp
i\varphi_D)\approx-2.2$, $\mathrm{Im}(h_F\exp i\varphi_F)\approx
6.1$, both in units of $\sqrt{2}F_\pi G_Fm_{\pi^+}^2J$. This leads
to:
\begin{equation}\varphi_D\approx
-1.5J,\varphi_F\approx-3.6J.\end{equation}

It is straightforward to see that $\tilde{\varphi}_D$ and
$\tilde{\varphi}_F$ receive no factorizable contribution. This is
because it would require terms like $\bar{\mathcal{R}}m_q B$ to
appear in the strong chiral Lagrangian. Such terms would violate
parity and therefore cannot exist. For the non-factorizable part, my
strategy is the following: first I compute the matrix elements
$\left\langle \mathcal{R}\right|\hat{Q}_6\left|B\right\rangle$ and
$\left\langle B'\right|\hat{Q}_6\left|B\right\rangle$ using the
quark model to find their ratio. Then, I use this ratio to infer the
value of the non-factorizable part of $\tilde{\varphi}_D$,
$\tilde{\varphi}_F$ by appropriately scaling the non-factorizable
part of $\varphi_D$, $\varphi_F$ given in Ref.
\cite{Tandean:2002vy}.

To obtain an estimate of hadronic matrix elements I adopt the
harmonic oscillator model \cite{LeYaouanc:1978ef}. The structure of
the spin-flavor wavefunction of the baryon octet leads to the
following ratio:
\begin{equation}\left\langle n^*\right|\hat{Q}_6\left|\Sigma^0\right\rangle:
\left\langle n^*\right|\hat{Q}_6\left|\Lambda\right\rangle:
\left\langle
p^*\right|\hat{Q}_6\left|\Sigma^+\right\rangle=1:\sqrt{3}:-\sqrt{2}\end{equation}
which requires that $w_F\tilde{\varphi}_F=(1/3)w_D\tilde{\varphi}_D$
in our chiral Lagrangian. I also obtain the ratio between $B-B'$ and
$B-\mathcal{R}$ matrix elements:
\begin{equation} \frac{\left\langle
p^*\right|\hat{Q}_6\left|\Sigma^+\right\rangle}{\left\langle
p\right|\hat{Q}_6\left|\Sigma^+\right\rangle}=-\sqrt{\frac{2}{3\pi}}\frac{1}{mR_0}.\end{equation}
where $m\approx 0.34\mathrm{GeV},R_0\approx 2.7\mathrm{GeV}^{-1}$
are harmonic oscillator parameters. With this ratio and the
non-factorizable contribution to $\varphi_D,\varphi_F$ given in
\cite{Tandean:2002vy}, I obtain the non-factorizable contribution to
$\tilde{\varphi}_D,\tilde{\varphi}_F$: \begin{equation}
\tilde{\varphi}_D\approx 0.04J,
\tilde{\varphi}_F\approx-0.01J\end{equation}These phases are about
two orders of magnitude smaller than the three other CP-phases
because they are not enhanced by the chiral quark condensate.
Therefore, I disregard them in the rest of our calculation.

To end this section, I point out that there is an important sign
issue in the determination of LECs. Since LECs are fitted to
experiments that only involve squared amplitudes, an overall
undetermined sign is left ambiguous. Therefore, if two sets of LECs
are fitted separately to two unrelated experiments (for example,
$\{r_D, r_F\}$ are to fit to baryon electromagnetic transitions and
$\{h_D,h_F,w_D,w_F\}$ are to fit to non-leptonic hyperon decays),
there is no unique way to determine the relative sign between these
two sets of LECs. This introduces an extra uncertainty because a
change of a relative sign can turn a constructive interference to
destructive and vice versa. I will discuss the impact of this
uncertainty in the last section.

\section{One loop contribution}

In this section I present analytic and numerical results of the
one-1oop contribution to the proton and neutron EDM using HBchPT.
The nucleon EDM $d_N$ is defined by the term linear in the incoming
photon momentum $q$ of the P and T-violating $NN\gamma$ amplitude

\begin{equation}iM\equiv-2d_N v\cdot\varepsilon\bar{u}_N S\cdot q u_N.\end{equation}
Here $\varepsilon^{\mu}$ is the photon polarization vector. Note
that the equation has been simplified by applying the on-shell
condition to the nucleon: $v\cdot q=-q^2/2m_N\rightarrow 0$.

Since each weak interaction vertex has $|\Delta S|=1$, I need at
least two insertions of weak interaction vertices to obtain an EDM
that is flavor diagonal. Most one-loop integrals are UV-divergent
and are regularized using the $\overline{\mathrm{MS}}$ scheme in
which the combination
\begin{equation}L\equiv\frac{2}{4-d}-\gamma+\ln(4\pi)\end{equation}is
subtracted. Also, since all CP-violating phases $\{\varphi_i\}$ are
small, I use the small angle approximation $\sin\varphi_i\approx
\varphi_i$. Finally, following the usual spirit of ChPT, during the
calculation of loops we assume that the heavy DOFs could be
integrated out and their effects show up in the LECs of the
effective operators consist of lighter DOFs \footnote{The reader
should anyway be alerted that this may not always be the case. For
example, Ref. \cite{Dashen:1993jt} pointed out that one needs to
include the baryon decuplet in order to reconcile with the result of
the large $N_c$-expansion}. Hence what enter the loops are the
lightest DOFs, which in our case are the pseudoscalar meson octet
and the ground-state $(1/2)^+$ baryon octet.

\begin{figure}
\includegraphics[scale=0.17]{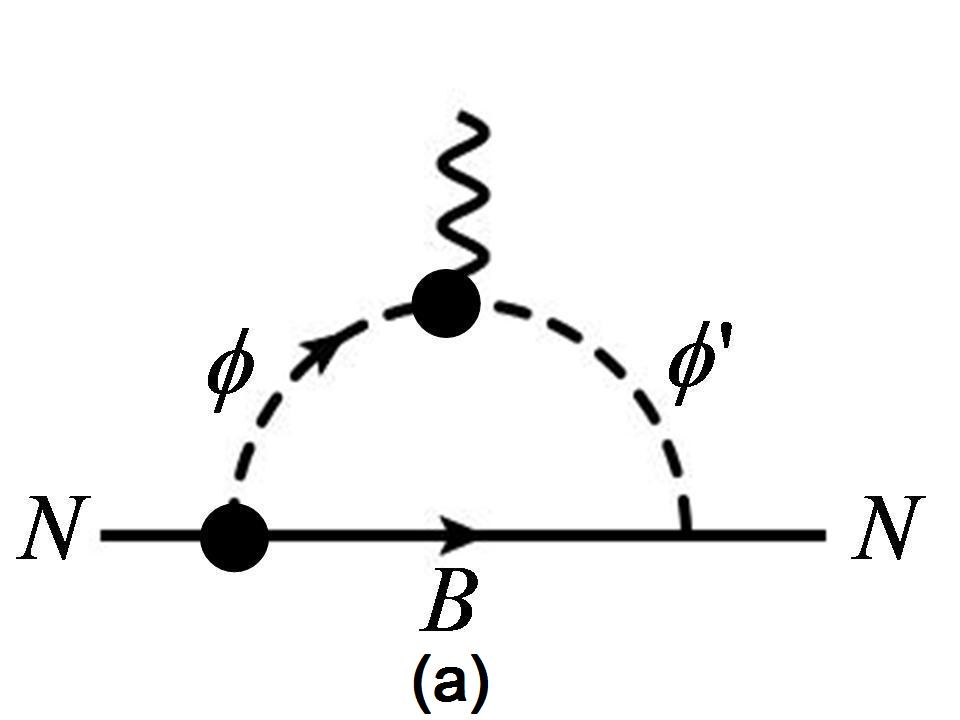}\includegraphics[scale=0.17]{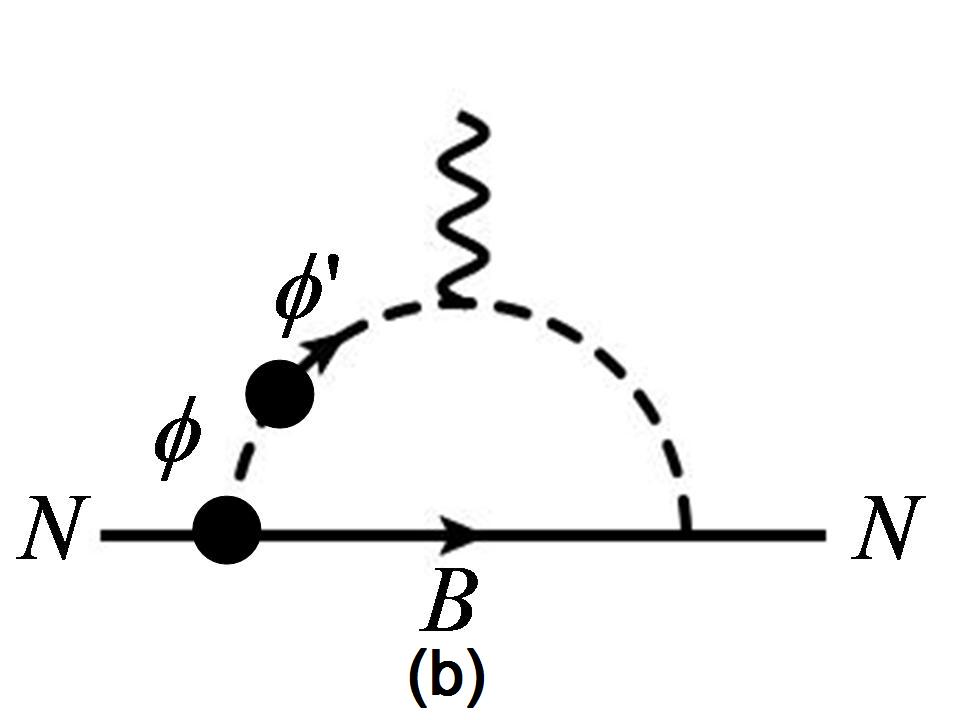}\includegraphics[scale=0.17]{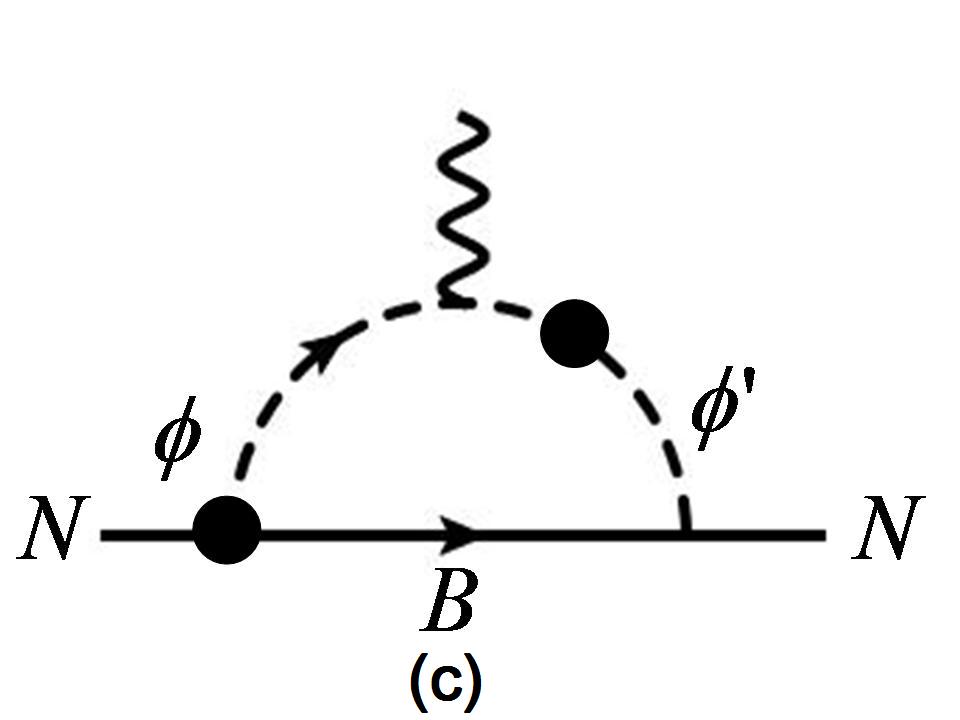}\includegraphics[scale=0.17]{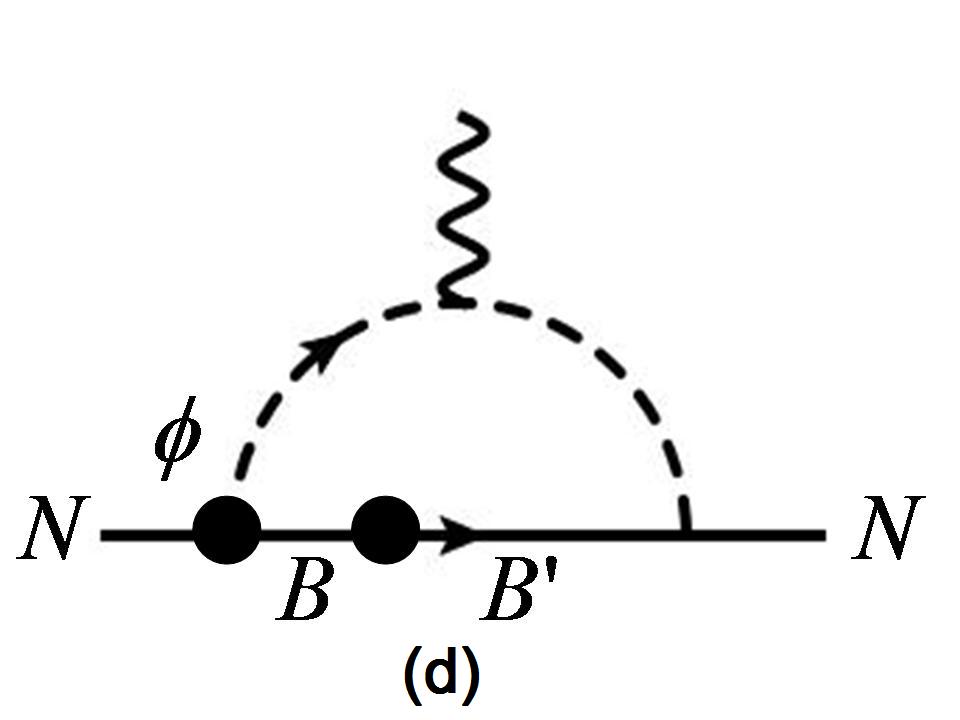}
\caption{\label{fig:1-loop}One-1oop contributions to the nucleon
EDM. Each round dot denotes a $|\Delta S|=1$ weak insertion. Fig.
1(a)-(c) (and reflections) contribute to both neutron and proton
EDM; while Fig. 1(d) (and reflection) contributes only to proton
EDM.}
\end{figure}

There are four distinct types of 1-loop diagrams (see
Fig.\ref{fig:1-loop}) that give non-zero contribution to the nucleon
EDM (diagrams of other kinds are all vanishing at leading order in
the HB-expansion. See the Appendix for the argument). Fig. 1(a)-(c)
(plus reflections) in Fig.\ref{fig:1-loop} contribute to both
neutron and proton EDM. For the neutron, it reads:
\begin{eqnarray}\label{eq:dn1}d_n^{\mathrm{1-loop}}&=&-\frac{eg_8(Dh_D\{\varphi-\varphi_D\}+Fh_F\{\varphi-\varphi_F\})}{4\pi^2F_\pi^4(m_\pi^2-m_K^2)}(m_\pi^2\ln\frac{m_\pi^2}{\mu^2}-\{\pi\leftrightarrow K\})\nonumber\\
&&-\frac{\delta_\Sigma
eg_8(D-F)(h_D\{\varphi-\varphi_D\}+h_F\{\varphi_F-\varphi\})}{4\pi^2F_\pi^2(m_\pi^2-m_K^2)}(m_\pi^2\frac{\arctan\frac{\sqrt{m_\pi^2-\delta_\Sigma^2}}{\delta_\Sigma}}{\sqrt{m_\pi^2-\delta_\Sigma^2}}-\{\pi\leftrightarrow
K\}).\nonumber\\
&&\end{eqnarray} I found that all terms analytic in quark masses
cancel each other. Also notice that there is no extra singularity in
the limit $m_K\rightarrow m_\pi$ or $\delta_B\rightarrow 0$.
Numerical estimation with $\mu=m_N$ gives
\begin{equation}|d_n^{\mathrm{1-loop}}|=1.5\times10^{-32}e\,\mathrm{cm}.\end{equation}

Similar calculations are done for the proton EDM. Figs. 1(a)-(c)
give
\begin{eqnarray}\label{eq:dp1}d_p^{\mathrm{1-loop,1}}&=&\frac{eg_8(D\{h_D[\varphi-\varphi_D]+3h_F[\varphi-\varphi_F]\}
+3F\{h_D[\varphi-\varphi_D]+h_F[\varphi_F-\varphi]\})}{24\pi^2F_\pi^4(m_\pi^2-m_K^2)}\times\nonumber\\
&&(m_\pi^2\ln\frac{m_\pi^2}{\mu^2}-\{\pi\leftrightarrow
K\})\nonumber\\
&&-\frac{\delta_\Sigma
eg_8(D-F)(h_D\{\varphi-\varphi_D\}+h_F\{\varphi_F-\varphi\})}{8\pi^2F_\pi^4(m_\pi^2-m_K^2)}(m_\pi^2\frac{\arctan\frac{\sqrt{m_\pi^2-\delta_\Sigma^2}}{\delta_\Sigma}}{\sqrt{m_\pi^2-\delta_\Sigma^2}}-\{\pi\leftrightarrow
K\})\nonumber\\
&&-\frac{\delta_\Lambda
eg_8(D+3F)(h_D\{\varphi-\varphi_D\}+3h_F\{\varphi-\varphi_F\})}{24\pi^2F_\pi^4(m_\pi^2-m_K^2)}(m_\pi^2\frac{\arctan\frac{\sqrt{m_\pi^2-\delta_\Lambda^2}}{\delta_\Lambda}}{\sqrt{m_\pi^2-\delta_\Lambda^2}}-\{\pi\leftrightarrow
K\}).\nonumber\\
&&\end{eqnarray}

There is one extra type of diagrams contributing to the proton EDM
corresponding to two insertions of $h_i$ vertices (Fig. 1(d)). The
corresponding diagrams do not generate the neutron EDM simply
because there is no appropriate non-vanishing combination of
$B,B',\phi$. This diagram for the proton EDM gives
\begin{eqnarray}\label{eq:dp2}d_p^{\mathrm{1-loop},2}&=&-\frac{eh_Dh_F(D-F)(\varphi_D-\varphi_F)(\pi-2\arctan\frac{\delta_\Sigma}{\sqrt{m_K^2-\delta_\Sigma^2}})}{16\pi^2F_\pi^2\sqrt{m_K^2-\delta_\Sigma^2}}\nonumber\\
&&-\frac{eh_Dh_F(D+3F)(\varphi_D-\varphi_F)(\pi-2\arctan\frac{\delta_\Lambda}{\sqrt{m_K^2-\delta_\Lambda^2}})}{48\pi^2F_\pi^2\sqrt{m_K^2-\delta_\Lambda^2}}.\end{eqnarray}
This contribution is interesting since it is UV-finite. It depends
non-analytically on quark masses and hence uniquely characterizes
long-distance physics \footnote{One can show that Eq. \eqref{eq:dp2}
remains real even when $\delta_K,\delta_\Lambda >m_K$ by using the
identity $\arctan z=\frac{1}{2i}\log\frac{1+iz}{1-iz}.$}.
Numerically, these give
\begin{eqnarray}|d_p^{\mathrm{1-loop,1}}|&=&6.1\times10^{-33}e\,\mathrm{cm}\nonumber\\
|d_p^{\mathrm{1-loop,2}}|&=&1.1\times10^{-32}e\,\mathrm{cm}.\end{eqnarray}
I choose to present numerical results of $d^{\mathrm{1-loop,1}}$ and
$d^{\mathrm{1-loop,2}}$ separately because the former is
proportional to $g_8 h_i$ while the latter is proportional to $h_i
h_j$. Since the relative sign between $g_8$ and $h_i$ is
experimentally undetermined, these two terms can either add or
subtract each other.

As a short conclusion, I stress once again that within the HBchPT
formalism, my analytic results of 1-loop diagrams, Eq.
\eqref{eq:dn1}, \eqref{eq:dp1} and \eqref{eq:dp2} fully respect
power counting as no powers of $m_B$ appear in the numerator upon
carrying out loop integrals. This is in contrast with the
relativistic calculation done in Ref. \cite{He:1989xj}, in which the
authors include diagrams involving MDM-like coupling that should
have an explicit $1/m_B$ suppression according to the power
counting, but is canceled by a factor of $m_B$ appearing in the
numerator coming from the loop integral.

Finally let me discuss the effect of counterterms. Since
$d_n^{\mathrm{1-loop}}$ and $d_p^{\mathrm{1-loop},1}$ are
UV-divergent, I need to introduce corresponding counterterms
${d_n^0,d_p^0}$ to absorb the infinities. These counterterms are
generated by short-distance physics. Therefore their precise values
cannot be calculated. To estimate the size of these counterterms we
perform a naive dimensional analysis (NDA). Following
\cite{Buchalla:1989we}, there are ten $\Delta S=1$ four-quark
operators that mix under renormalization. The effective Hamiltonian
can be written as:
\begin{equation}H^{\Delta
S=1}_{\mathrm{eff}}=\frac{G_F}{\sqrt{2}}V_{\mathrm{ud}}V_{\mathrm{us}}^*\sum_{i=1}^{10}
C_i(\mu)\hat{Q}_i(\mu)+h.c.\end{equation} Under conditions that
$\Lambda_{\mathrm{QCD}}\approx0.2\mathrm{GeV}$, $\mu=1\mathrm{GeV}$
and the top-quark mass $m_t=174\mathrm{GeV}$, the largest
flavor-diagonal CP-violating effect comes from the product of
$\hat{Q}_2$ and $\hat{Q}_6$ with Wilson coefficients
$C_2=1.31-0.044\tau$ and $C_6=-0.011-0.080\tau$ where
$\tau=-V_{\mathrm{td}}V_{\mathrm{ts}}^*/V_{\mathrm{ud}}V_{\mathrm{us}}^*$.
This gives:
\begin{equation}\label{eq:NDA}d_p^0,d_n^0\sim\frac{1}{16\pi^2}\frac{G_F^2}{2}|V_{\mathrm{ud}}V_{\mathrm{us}}^*|^2\mathrm{Im}(C_2C_6^*)\Lambda_\chi^3\approx4\times10^{-32}e\,\mathrm{cm}.\end{equation}
Here $1/16\pi^2$ is a necessary loop factor while the factor
$\Lambda_\chi^3$ is included to achieve the correct mass dimension.
I choose $\Lambda_\chi\sim1\mathrm{GeV}$ instead of some other scale
like $\Lambda_{\mathrm{QCD}}\sim200\mathrm{MeV}$ to provide a
conservative upper limit for $d_p^0$ and $d_N^0$. This analysis
shows that the short-distance contribution to the nucleon EDM could
be as large as the long-distance contribution\footnote{A follow-up
work from the author to compute these short-distance contributions
within certain nucleon model framework is currently in progress.}.
However the NDA estimation is rarely trustable and it may happen
that some accidental cancelations could suppress the actual value of
$d_n^0,d_p^0$ from what is expected in Eq. \eqref{eq:NDA}. In this
sense, a detailed study of the long-distance contribution is
worthwhile because it sets a solid bound below which any measurable
nucleon EDM could be safely regarded as being consistent with the SM
prediction.

\section{Pole Contribution}

\begin{figure}
\includegraphics[scale=0.15]{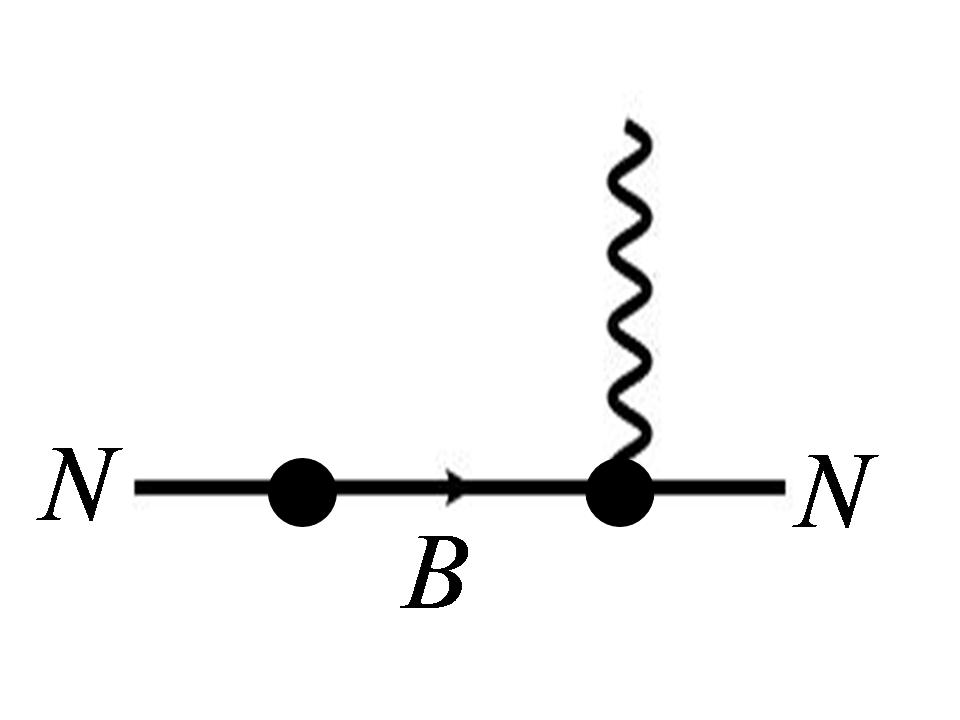}\includegraphics[scale=0.15]{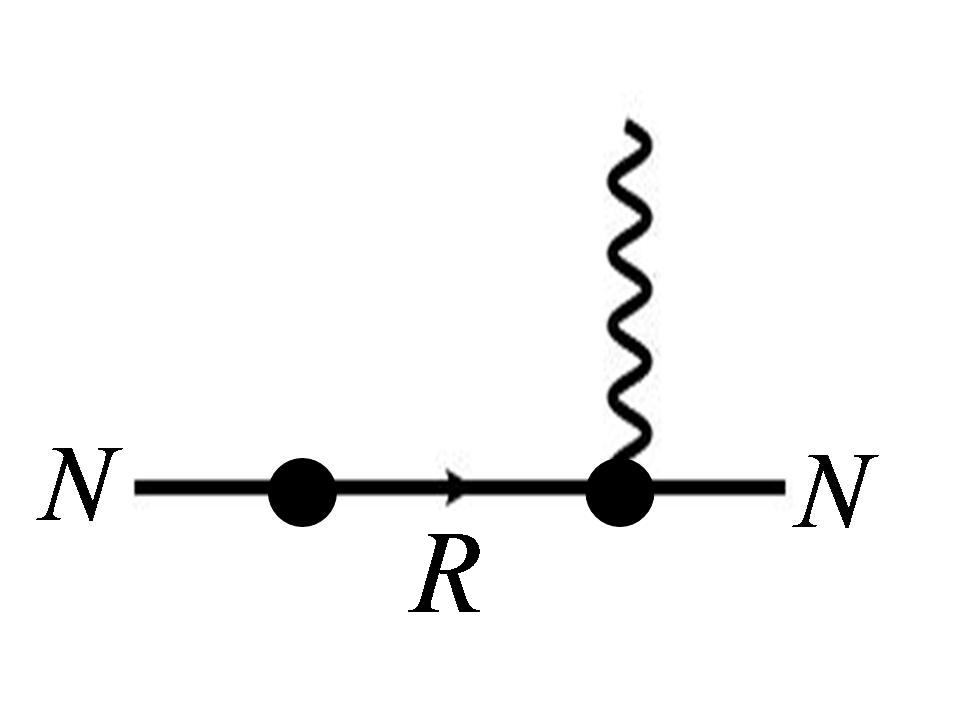}\caption{\label{fig:classI}(with reflections) Class
I pole diagrams.}\end{figure}

\begin{figure}
\includegraphics[scale=0.15]{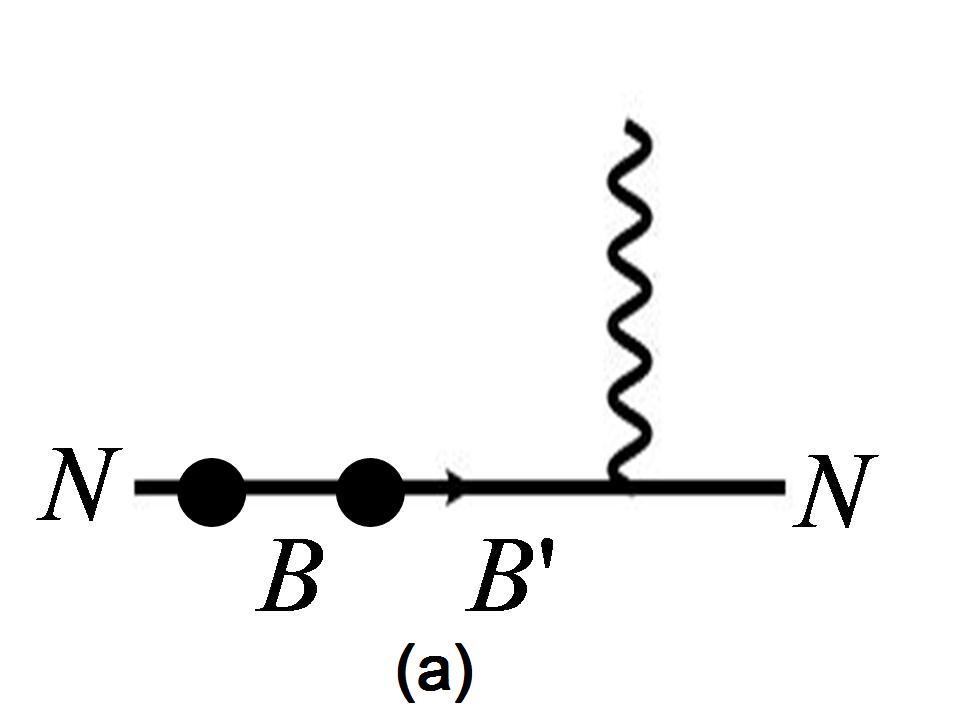}\includegraphics[scale=0.15]{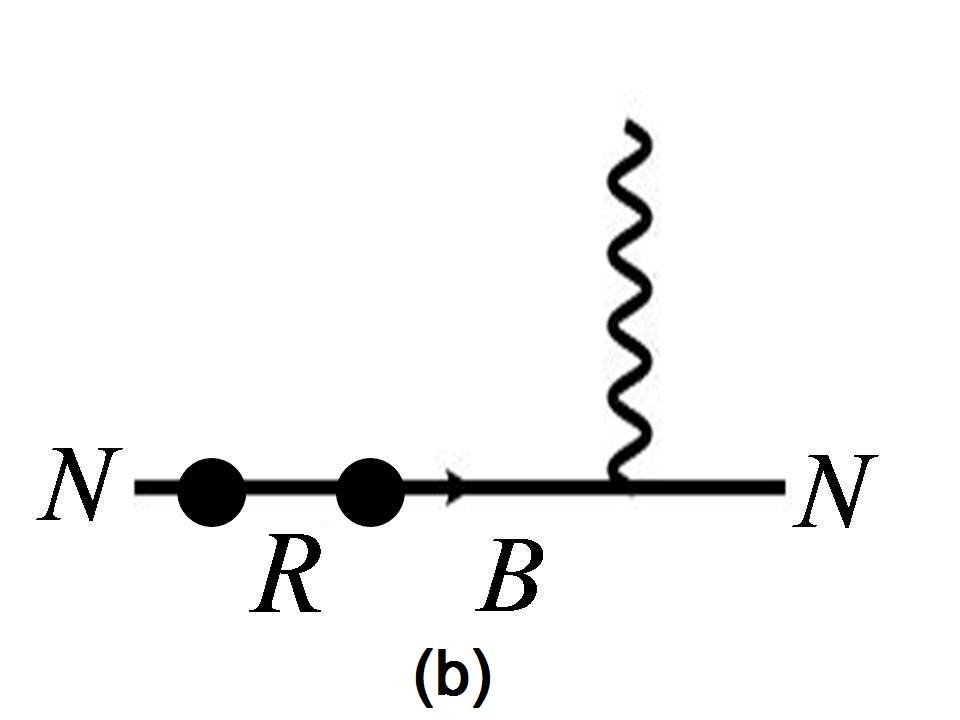}\includegraphics[scale=0.15]{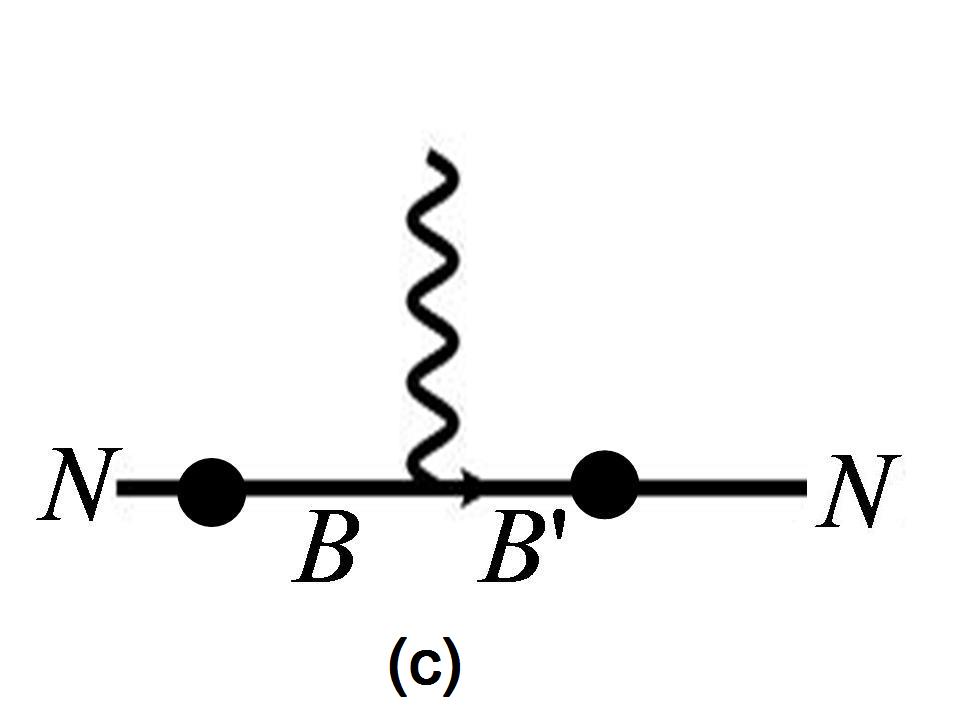}\includegraphics[scale=0.15]{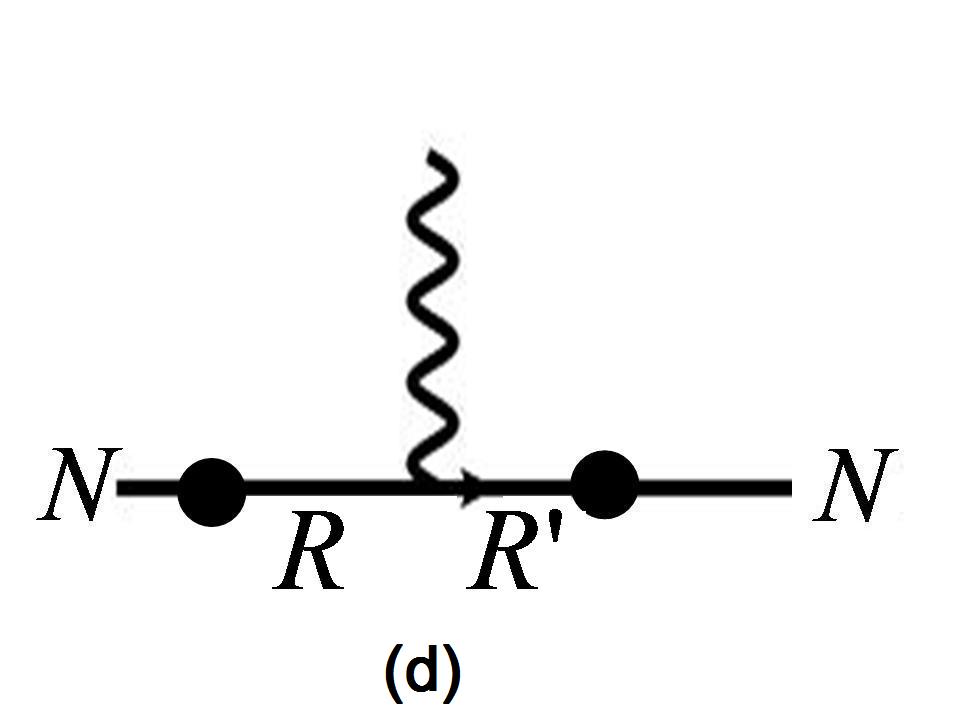}
\includegraphics[scale=0.15]{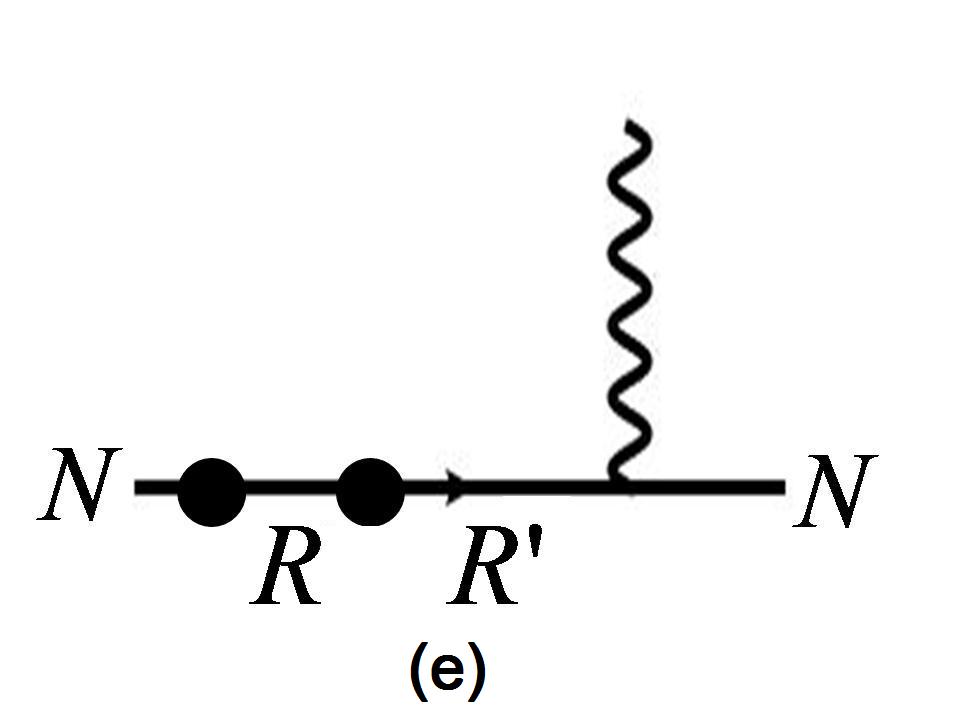}\includegraphics[scale=0.15]{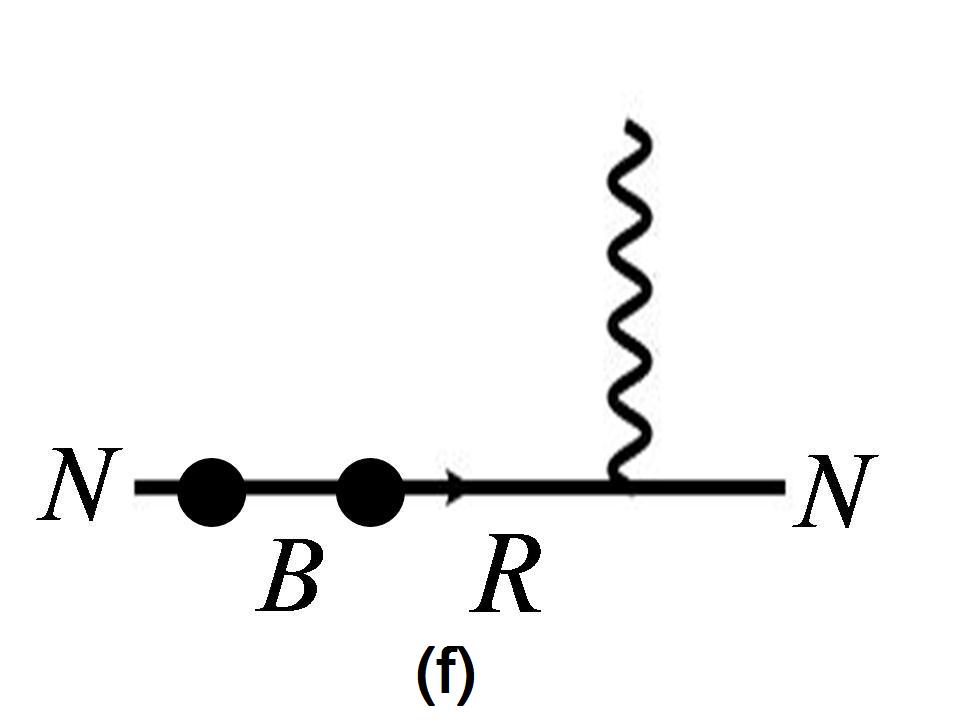}\includegraphics[scale=0.15]{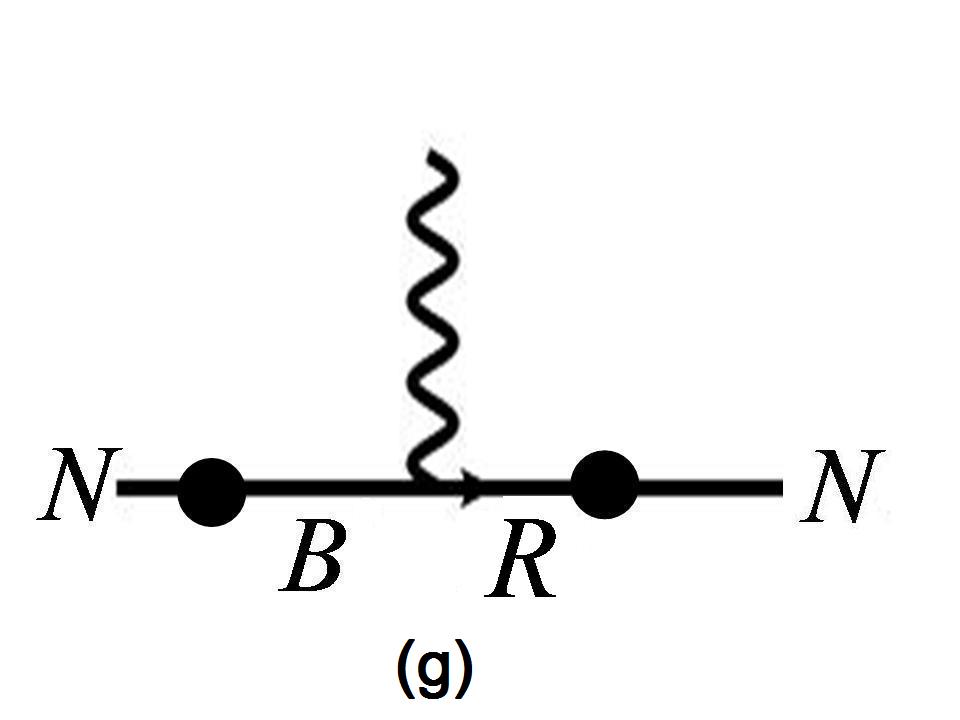}\caption{\label{fig:classII}(with reflections) Class
II pole diagrams.}\end{figure}

Next I estimate the contribution of pole diagrams to the nucleon
EDM. For baryon intermediate states, I include the flavor octet part
of the $(56,0^+)$ and $(70,1^-)$ baryon supermultiplets. Here I
adopt the standard spin-flavor $SU(6)$ notation $(\mathcal{D},L^p)$
where $\mathcal{D}$ is the dimension of the $SU(6)$ representation,
$L$ is the orbital angular momentum and $p$ is the parity. For
generality, we first write down all possible pole configurations
that can contribute and divide it into two classes: Class I are
those in which the photon vertex involves a weak insertion and Class
II are those in which the photon vertex is purely electromagnetic
(see Fig \ref{fig:classI} and \ref{fig:classII}).

I want to single out the leading pole diagrams. First, one would
expect that Class I contributions are much smaller than Class II for
two reasons: (1) the weak photon vertex in Class I diagrams is due
to the transition quark magnetic dipole moment (MDM) that contains a
$m_s+m_d$ suppression factor or the transition quark EDM that is
suppressed by $m_s-m_d$ (the latter, which vanishes if
$m_s\rightarrow m_d$, is an explicit demonstration of Hara's theorem
\cite{Hara:1964zz}); (2) Class II diagrams have one more pole in the
denominator. With these observations I may safely discard Class I
diagrams since they are sub-leading.

Within Class II, Fig. 3(a)-(d) can be shown to have an extra $1/m_N$
suppression \cite{Seng:2014pba}. These four diagrams involve
MDM-like baryon radiative transition vertices that have the
structure of $(1/m_B)\epsilon^{\mu\nu\alpha\beta}v_\nu q_\alpha
S_\beta$ at leading order. This structure is orthogonal to the EDM
structure $v^\mu S\cdot q$ so it cannot generate an EDM. Therefore
in order to obtain an EDM one needs to go to the next order in the
HB-expansion leading to an extra $1/m_N$ suppression, so I can
discard these four diagrams. Finally, Fig. 3(e) is smaller than Fig.
3(f)-(g) due to an extra propagator of a heavy excited state $R$.
After all these considerations, I only need to evaluate Fig.
3(f)-(g). Using Feynman rules obtained from the Lagrangian in
Section II, I obtain
\begin{eqnarray}\label{eq:pole}
d_n^{\mathrm{pole}}&=&\frac{4er_D}{9\delta_\Lambda\delta_{\Lambda^*}\delta_{N^*}\delta_{\Sigma^*}\delta_\Sigma}(h_D\varphi_D\{3w_F[2\delta_{\Lambda^*}\delta_{\Sigma^*}(\delta_\Lambda-\delta_\Sigma)
+\delta_{N^*}\{\delta_{\Lambda^*}(\delta_\Lambda+\delta_\Sigma)\nonumber\\
&&+\delta_{\Sigma^*}(\delta_\Sigma-3\delta_\Lambda)\}]-w_D[2\delta_{\Lambda^*}\delta_{\Sigma^*}(3\delta_\Lambda+\delta_\Sigma)+\delta_{N^*}\{3\delta_{\Lambda^*}(\delta_\Lambda+\delta_\Sigma)\nonumber\\
&&+\delta_{\Sigma^*}(3\delta_\Lambda-\delta_\Sigma)\}]\}+3h_F\varphi_F\{w_D[2\delta_{\Lambda^*}\delta_{\Sigma^*}(\delta_\Lambda-\delta_\Sigma)+\delta_{N^*}\{\delta_{\Lambda^*}(\delta_\Lambda-3\delta_\Sigma)\nonumber\\
&&+\delta_{\Sigma^*}(\delta_\Lambda+\delta_\Sigma)\}]+w_F[\delta_{N^*}\{3\delta_{\Sigma^*}(\delta_\Lambda+\delta_\Sigma)-\delta_{\Lambda^*}(
\delta_\Lambda-3\delta_\Sigma)\}\nonumber\\
&&-2\delta_{\Lambda^*}\delta_{\Sigma^*}(\delta_\Lambda+3\delta_\Sigma)]\})\nonumber\\
d_p^{\mathrm{pole}}&=&-\frac{8e(\delta_{N^*}-\delta_{\Sigma^*})(r_D+3r_F)(w_D-w_F)(h_D\varphi_D-h_F\varphi_F)}{3\delta_{N^*}\delta_{\Sigma^*}\delta_\Sigma}.\end{eqnarray}
In the expression above I have neglected the two small phases
$\tilde{\varphi}_D$ and $\tilde{\varphi}_F$. Note that Eq.
\eqref{eq:pole} diverges in the $\delta\rightarrow 0$ limit. This
simply indicates that non-degenerate perturbation theory fails in
this limit and one needs to switch to degenerate perturbation
theory. Numerically, Eq. \eqref{eq:pole} gives:
\begin{equation}
|d_n^{\mathrm{pole}}|\approx|d_p^{\mathrm{pole}}|\approx 1.4\times
10^{-32}e\,\mathrm{cm}.\end{equation}

\begin{table}
\begin{centering}
\begin{tabular}{|c|c|c|c|}
\hline
 Nucleon\textbackslash{}EDM &$|d_N^{\mathrm{1-loop},1}|$&$|d_N^{\mathrm{1-loop},2}|$&$|d_N^{\mathrm{pole}}|$ \tabularnewline
\hline \hline neutron & $1.5\times10^{-32}$ & 0 &
$1.4\times10^{-32}$\tabularnewline \hline proton &
$6.1\times10^{-33}$ & $1.1\times10^{-32}$ &
$1.4\times10^{-32}$\tabularnewline \hline
\end{tabular}
\par\end{centering}

\caption{\label{tab:table}Different contributions to the SM neutron
and proton EDM in units of $e\,\mathrm{cm}$, assuming the sign of
LECs are those given in Section III.}\end{table}

Numerical results are summarized in Table \ref{tab:table}. I caution
the readers that all these numbers are only indicative of the size,
because I have not yet addressed the sign ambiguities plaguing the
determination of certain LECs as emphasized at the end of section
III. This will be done in the next section.

\section{Discussion and Summary}

Now I consider the uncertainty due to the undetermined relative sign
between different groups of LECs. Since $r_D$ and $r_F$ are fitted
simultaneously to the electromagnetic decay of $(1/2)^-$ resonance
they should be multiplied by a common undetermined sign factor
$\delta_r=\pm1$. The constant $g_8$ is fitted to the kaon decay
rate, so it should carry a separate sign factor $\delta_g$. Its
phase $\varphi$ however is determined theoretically so it does not
have a sign ambiguity. The four remaining interaction strengths
$\{h_D,h_F,w_D,w_F\}$ are fitted simultaneously to s and p-wave
amplitudes of the hyperon non-leptonic decay, so they should carry a
common undetermined sign factor $\delta_{hw}$. Their corresponding
phases are determined by first calculating
$\mathrm{Im}\{h_i\exp{i\varphi_i}\}$ and
$\mathrm{Im}\{w_i\exp{i\tilde{\varphi}_i}\}$ theoretically and then
by dividing them by the experimentally-determined $\{h_i,w_i\}$ so
the four remaining phases
$\{\varphi_D,\varphi_F,\tilde{\varphi}_D,\tilde{\varphi}_F\}$ should
also carry the same sign factor $\delta_{hw}$. Summing up loop and
pole diagram contributions and allowing
$\{\delta_r,\delta_g,\delta_{hw}\}$ to freely change between 1 and
-1, I obtain a range of possible $d_n$ and $d_p$:
\begin{eqnarray}&&8.7\times10^{-34}e\,\mathrm{cm}<|d_n|<2.8\times10^{-32}e\,\mathrm{cm}\nonumber\\
&&3.3\times10^{-33}e\,\mathrm{cm}<|d_p|<3.3\times10^{-32}e\,\mathrm{cm}\end{eqnarray}

The surprisingly small lower bounds of $|d_n|,|d_p|$ are due to an
accidental cancelation between loop and pole-diagram contributions
for a very specific set of $\{\delta_i\}$. There is no reason to
believe that this cancellation persists at higher order. To estimate
the size of higher-order contributions, I recall that the
HB-expansion parameter is of order $m_K/m_N\sim0.5$. Therefore to be
conservative, I could assign a 100\% error due to the
next-to-leading-order (NLO) effects in the HB-expansion. Also, by
looking at Table \ref{tab:table} one can see that both loop and pole
diagrams are of order $10^{-32}e\,\mathrm{cm}$. So if I assume no
fine cancellation between these two parts after adding the NLO
contributions from the HB-expansion, then I should expect the
long-distance contribution to the nucleon EDM to lie within the
range:
\begin{equation}\label{eq:final} 1\times10^{-32}e\,\mathrm{cm}<\{|d_n|,|d_p|\}<6\times10^{-32}e\,\mathrm{cm}.\end{equation} My estimated upper bound for $d_n$ is about half the corresponding value
predicted in \cite{He:1989xj}. Eq. \eqref{eq:final} is three (four)
orders of magnitude smaller than the proposed precision level of the
future proton (neutron) EDM experiments.

To summarize, even though it is well-known that the nucleon EDM
induced by the Standard Model CKM matrix is well below the limit of
our current experimental precision, it is still worth a thorough
study as it is currently the only source of intrinsic EDMs in nature
whose existence is certain. I re-analyze previous works on chiral
loop and pole diagram contributions to the nucleon EDM using HBchPT
at the leading order in HB-expansion, with an up-to-date
determination of relevant LECs that enter our calculation. Combined
with the uncertainty due to unknown relative signs of LECs and an
estimate of higher-order contributions, I obtain the range for the
long-distance contribution to the nucleon EDM in Eq.
\eqref{eq:final}. Although an incalculable short-distance physics
which appears as counterterms in our work could be as large as the
long-distance contribution, the study of the long-distance
contribution is still worthwhile as it provides a safe borderline
below which any nucleon EDM is consistent with the SM prediction.
Finally, there are several ways to improve upon the estimate carried
out in this work. For instance, a combined analysis of lattice
simulations and better experimental measurements of various hadronic
decay processes is expected to provide a better control of both the
magnitudes and signs of the required LECs. If the LECs could be
determined more precisely, then a complete analysis of NLO-effects
in the HB-expansion would be much desired to further restrict the
allowed range of $d_n$ and $d_p$.

\section{Acknowledgements}

The author would like to thank Michael J. Ramsey-Musolf, John F.
Donoghue and Barry R. Holstein for many useful discussions. The
author is also grateful to Hiren H. Patel and Graham White for
carefully reading the manuscript and providing extensive valuable
feedback. This work is supported in part by the US Department of
Energy under contract No. DE-SC0011095.

\appendix

\section{Vanishing one-loop diagrams}

Here I will show that all 1-loop diagrams, other than those in Fig.
\ref{fig:1-loop}, do not give rise to the nucleon EDM, at least at
leading order in the HB-expansion.

\begin{figure}
\includegraphics[scale=0.17]{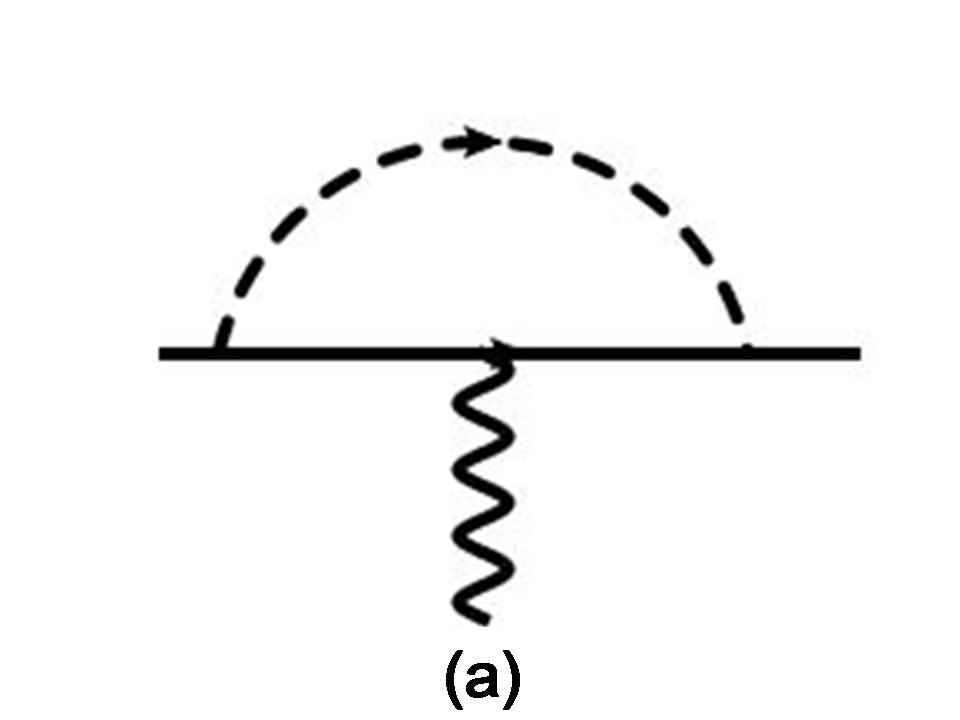}\includegraphics[scale=0.17]{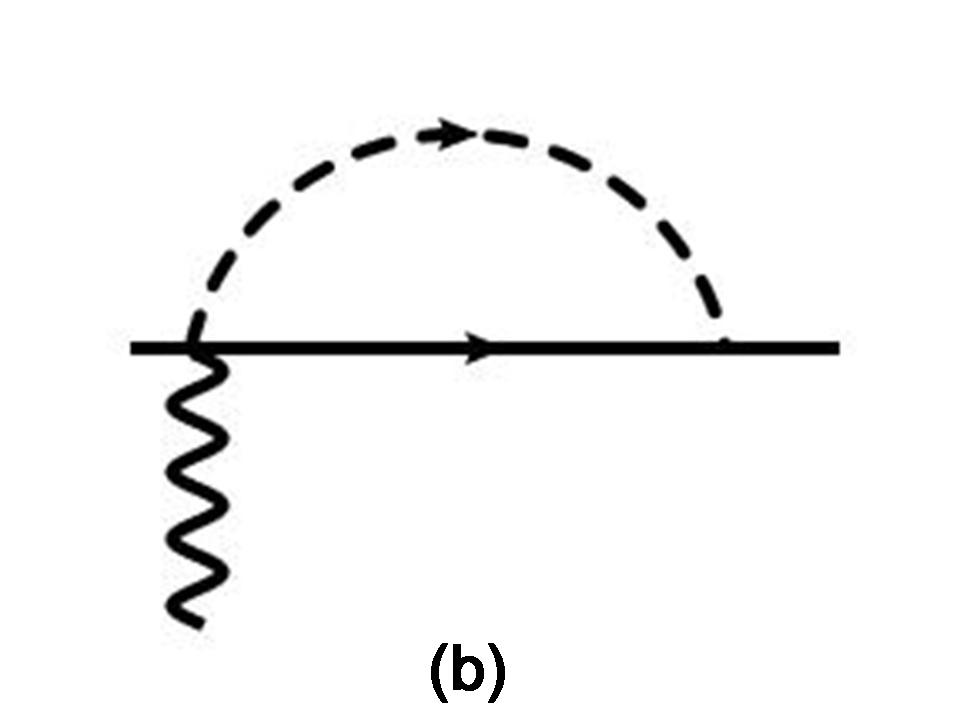}\includegraphics[scale=0.17]{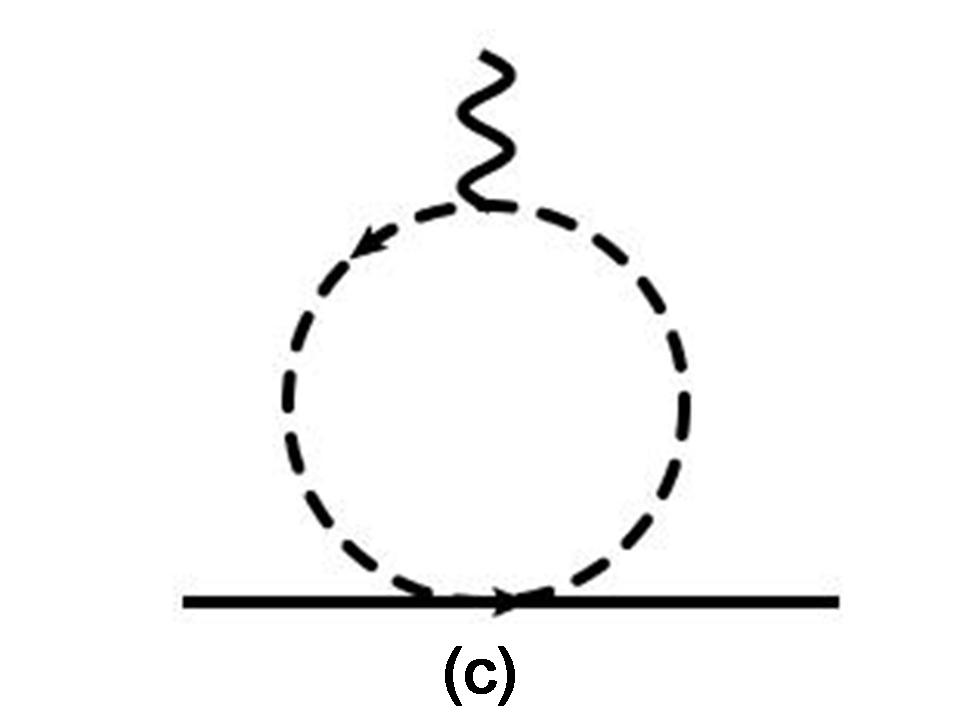}\includegraphics[scale=0.17]{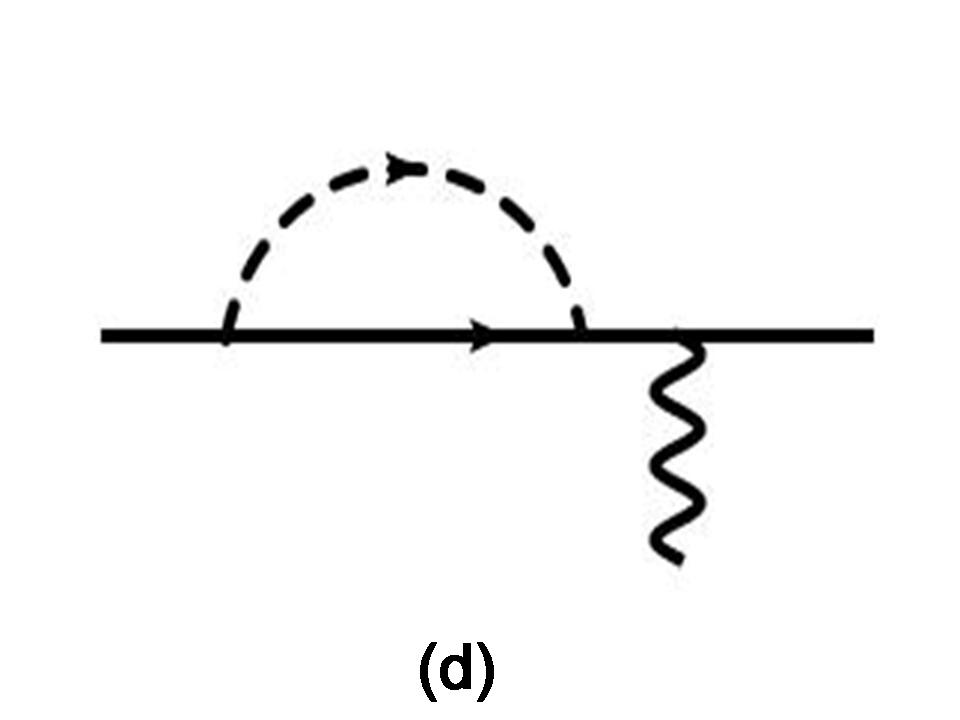}
\caption{\label{fig:vanish}1-loop diagrams that vanish at LO HBchPT.
The weak vertices could be placed at any allowed position and
therefore are not explicitly shown.}
\end{figure}

All other possible 1-loop diagrams beside those I have calculated
are summarized in Fig. \ref{fig:vanish}. Since the weak Lagrangian
used in my work does not involve covariant derivatives of baryon
fields, any baryon-photon coupling term has to arise from the
ordinary P and T-conserving Lagrangian.

For Fig. 4(a), the photon vertex must arise from Dirac coupling
since an MDM coupling is suppressed by $(1/m_N)^2$ as pointed out in
\cite{Seng:2014pba}. Since the Dirac coupling is independent of the
photon momentum $q$, one can define loop momenta in a way such that
the dependence of $q$ only appears in the baryon propagator.
However, using the on-shell condition $v\cdot q=0$, the baryon
propagator is actually $q$-independent and therefore so is the whole
diagram. As a result, Fig. 4(a) cannot generate an EDM that is
linear in $q$.

For Fig. 4(b), at leading order in the HB-expansion the
$BB'\phi\gamma$ vertex is proportional to $S^\mu$, so it cannot
generate an EDM because the latter is proportional to $v^\mu$ which
is perpendicular to $S^\mu$.

For Fig. 4(c), first I note that the $BB'\phi\phi'$ vertex cannot
come from the $D$ {\color{red}or} $F$-term of the ordinary chiral
Lagrangian because that would violate parity. Therefore it can only
come from $\mathcal{L}_w^{(s)}$. In this case, it can only be
parity-conserving and time reversal-conserving (PCTC), or
parity-conserving and time reversal-violating (PCTV). So in order to
get an EDM which is PVTV, one needs to place another PVTC or PVTV
vertex in some other part of the diagram. This cannot be done
because all $\phi\phi'$ and $\phi\phi'\gamma$ operators I have are
parity-conserving.

For Fig. 4(d), one could generate an EDM by coupling the resulting
complex mass term of the baryon to its MDM. But again this
contribution is suppressed by $(1/m_N)^2$ and should be discarded at
leading order in the HB-expansion.

\end{document}